\documentclass[prd,a4paper,preprint,showpacs,byrevtex,onecolumn]
{revtex4}
\usepackage{graphicx}
\usepackage{dcolumn}
\usepackage{amsmath}
\usepackage{array}
\usepackage{bm}
\usepackage{textcomp}

\DeclareMathAlphabet{\mathbbm}{U}{bbm}{m}{n}
\SetMathAlphabet\mathbbm{bold}{U}{bbm}{bx}{n}
\begin{document}

\title{Mass formula for strange baryons in large $N_c$ QCD versus quark
model}
\author{Claude \surname{Semay $^{a}$}}
\email[E-mail: ]{claude.semay@umh.ac.be}
\author{Fabien \surname{Buisseret $^{a}$}}
\email[E-mail: ]{fabien.buisseret@umh.ac.be}
\author{Florica \surname{Stancu} $^{b}$}
\email[E-mail: ]{fstancu@ulg.ac.be}
\affiliation{$^{a}$ Groupe de Physique Nucl\'{e}aire Th\'{e}orique,
Universit\'{e} de Mons-Hainaut,
Acad\'{e}mie universitaire Wallonie-Bruxelles,
Place du Parc 20, B-7000 Mons, Belgium.\\
$^{b}$ University of Li\`ege, Institute of Physics B5, Sart Tilman,
B-4000 Li\`ege 1, Belgium.}

\date{\today}

\begin{abstract}
A previous work establishing a connection between a quark model, with
relativistic
kinematics and a $Y$-confinement plus one gluon exchange,
and the $1/N_c$ expansion mass formula is extended
to strange baryons. Both methods predict values for the SU(3)-breaking
mass
terms which are in good agreement with each other. Strange and
nonstrange baryons are shown
to exhibit Regge trajectories with an equal slope, but with an intercept
depending on the strangeness. Both approaches agree on the value of the
slope and
of the intercept and on the existence of a single good quantum number
labeling the baryons within a given Regge trajectory.
\end{abstract}




\maketitle
\section{Introduction}

In Ref.~\cite{lnc}
we have made a first attempt to establish a connection
between quark model results for baryon masses and the
$1/N_c$ expansion mass formula. So far we have considered
nonstrange baryons only. The purpose of the present
study is to extend the previous work to strange baryons.

The standard approach to baryon spectroscopy is
the constituent quark model where the results are model dependent.
The states are classified according to SU(6) symmetry.
The phenomenological analysis suggested that
the baryons can be grouped into excitation bands $N$ = 0,1,2,...
each band containing at least one SU(6) multiplet.
The hyperfine interaction breaks the SU(6) symmetry.
The introduction of $N$ as a good quantum number for a Hamiltonian
with a linear $Y$-junction confinement is quite natural for
nonstrange baryons \cite{lnc}. We shall
show that, even for strange baryons, $N$ is a good
classification number within the same quark model.
The problem has already been discussed qualitatively in 
a nonrelativistic model with a quadratic two-body confinement \cite{Isgur}. 
Here we provide a quantitative 
proof of the role of the strange quark mass $m_s \neq m_{u,d}$ within a
semi-relativistic model with a realistic confinement.

The $1/N_c$ expansion method \cite{HOOFT,WITTEN}
offers an alternative, model independent
way, to study baryon spectroscopy in a systematic way.
The method stems from the discovery that
in the limit $N_c \rightarrow \infty$,
where $N_c$ is the number of colors, QCD possesses an exact contracted
SU(2$N_f$) symmetry \cite{Gervais:1983wq,DM} where $N_f$ is the number
of flavors. This symmetry is
only approximate for finite $N_c$ so that corrections have to be added
in powers of $1/N_c$. Here we discuss  the case $N_f$ = 3.
Thus, SU(6) is a common symmetry for both approaches.
The $1/N_c$ expansion method has extensively and successfully been
applied
to ground state  baryons ($N$ = 0)
\cite{DJM94,DJM95,Jenkins:1998wy,EJ2}.
Its applicability to excited states is a subject of current
investigations.  The most studied bands
so far are $N$ = 1 and 2.

It is important to compare the two methods. This could bring support
to quark model assumptions on one hand and it could help to gain more
physical insight into the dynamical coefficients of the $1/N_c$
expansion mass formula on the other hand. The key aspect in this 
comparison is that
one can analyze both the $1/N_c$ expansion and quark model results
in terms of $N$.
The paper is organized as follows. The mass formula used in the $1/N_c$
expansion for strange baryons is introduced in the next section.
Section~\ref{qmsb} gives a corresponding mass formula obtained from a
Hamiltonian quark model of spinless Salpeter type, where the confinement
is a $Y$-junction flux tubes and where the Coulomb-type one gluon
exchange
and the quark
self-energy contributions are added perturbatively.
There, we analytically prove that the classification of light baryons,
containing $u,d,s$
quarks, is still possible in terms of a quantum number $N$, representing
units of excitation, like in a harmonic oscillator picture.
Section~\ref{compar}
is devoted to the comparison of the results derived from  the $1/N_c$
expansion on one hand and from the quark model on the other hand.
A discussion on Regge trajectories emerging from the quark model
mass formula is also given. Conclusions
are finally drawn in Sec.~\ref{conclu}.


\section{Strange baryons in large $N_c$ QCD}\label{s_bar}
\subsection{Mass formula}\label{mass_f}

For strange baryons the mass operator in the
$1/N_c$ expansion has the general form
\begin{equation}
\label{massoperator}
M = \sum_{i=1} c_i O_i + \sum_{i=1} d_i B_i,
\end{equation}
where the operators $O_i$ are invariants under SU(6) transformations
and the operators $B_i$ explicitly break SU(3)-flavor symmetry.
The coefficients $c_i$ and $d_i$ are fitted
from the experimental data and encode the quark dynamics. In the case
of nonstrange baryons, only the operators $O_i$ contribute while
$B_i$ are defined such as their expectation values are zero.
Thus in  Ref.~\cite{lnc}, devoted to nonstrange baryons, only the
first term of Eq.~(\ref{massoperator}) entered the discussion.
Presently we focus
on the second term. The mass $m_s$ of the strange
quark breaks SU(3)-flavor explicitly.  The operators $B_i$ and
coefficients
$d_i$ are used to construct a mass shift $\Delta M_s$ introduced below.

In Eq.~(\ref{massoperator}) the sum over $i$ is finite and in practice
it includes the most dominant operators.
The building blocks of $O_i$ and $B_i$ are the SU(6) generators: $S_i$
($i$ = 1,2,3) acting on spin and forming an su(2) subalgebra,
$T^a$ ($a$ = 1,...,8)
acting on flavor and forming an su(3) subalgebra, and $G^{ia}$ acting
both on spin and flavor subspaces. For orbitally excited states, also
the components $\ell_i$ of the angular momentum,
as generators of SO(3), and the tensor operator $\ell^{ij}$
(see \emph{e.g.} Ref.~\cite{Matagne:2006zf})
are necessary to build $O_i$ and $B_i$.
Examples of $O_i$ and $B_i$ can be found in
Refs.~\cite{Goity:2002pu,Goity:2003ab,Matagne:2004pm,Matagne:2006zf}.
Each operator  $O_i$ or $B_i$ carries an explicit factor of
$1/N^{n-1}_c$
resulting from the power counting rules \cite{WITTEN}, where $n$
represents the minimum of gluon exchanges to generate the operator.
In the matrix elements there are also compensating factors of $N_c$
when one sums coherently over $N_c$ quark lines. In practice
it is customary to drop higher order corrections of order $1/N^2_c$.

We assume that each strange quark brings
the same contribution $\Delta M_s$ to the SU(3)-breaking
terms in the mass formula. To make a comparison between
the $1/N_c$ expansion and the quark model results we define
$\Delta M_s$ to satisfy the relation
\begin{equation}\label{break}
n_s ~\Delta M_s = \sum_{i=1} d_i B_i
\end{equation}
where $n_s $ is the number of strange quarks in a baryon.

Previous studies have indicated that, to a very
good approximation, one can apply the $1/N_c$ expansion mass formula to a
specific band and neglect interband mixing (see \emph{e.g.}
Refs.~\cite{Goity:2002pu,Goity:2003ab,Matagne:2004pm,Matagne:2006zf}),
so that we can make a comparison of the results
of both approches, band by band.

For $N$ = 0, 1, and 3, we adopt the values of $\Delta M_s$ provided
by Ref.~\cite{GM} and exhibited below in Table~\ref{tabdm}. Actually,
for $N$ = 3, the only
available values are from this reference, where they have been
calculated in an approximate way. The large error bars
suggest that this band must be more precisely reanalyzed in the large
$N_c$
approach. The fit has been made on the $[70,3^-]$ multiplet, which is
the lowest
among the eight multiplets contained in this band \cite{SS91}.

For $N$ = 4, Table~I of Ref.~\cite{Matagne:2004pm} straightforwardly
gives
$\left. \Delta M_s \right|_{N=4} = b_1 = 110 \pm 67$~MeV. Due to the
scarcity of data, there the analysis
was restricted to $[56,4^+]$, also the lowest SU(6) multiplet, out of
17 theoretically found multiplets,
in this band \cite{SS94}.

For $N$ = 2 the data are richer. In the next subsection we show how to
estimate
the mass shift  $\Delta M_s$ for $N$ = 2, by using values of
$d_i$ determined in previous $1/N_c$ expansion studies of several
SU(6) multiplets.

\subsection{The SU(3)-breaking in the $N =2$ band}\label{lnc_s}

Here we discuss details of the SU(3)-breaking for orbitally excited
baryons belonging to
the $[56, 2^+]$, $[70,0^+]$ or $[70,2^+]$ multiplets of the $N = 2$
band. The $[20, 1^+]$ multiplet of the
$N = 2$ band is not physically relevant. We combine the large $N_c$
results obtained for
$[56, 2^+]$ in Ref.~\cite{Goity:2003ab} and for $[70,\ell^+]$ 
($\ell= 0, 2$)
in Ref.~\cite{Matagne:2006zf}.
For the $[56, 2^+]$ multiplet the situation is simple.
There are 18 strange resonances in this sector.
The analysis of Ref.~\cite{Goity:2003ab} gives
$\left. \Delta M_s \right|_{[56,2^+]} = 206 \pm 18$~MeV.

For $[70, \ell^+]$ the situation is more complicated. The SU(3)-breaking
can be measured  by means of
Eq.~(\ref{break}) where in the right hand side we replace $B_i$ by
their expectation values. In this case there are two dominant
operators, $B_1$ and $B_2$, and we have $d_1 = 365 \pm 169$~MeV and
$d_2 = -293\pm54$~MeV from Ref.~\cite{Matagne:2006zf}.
Then for a given baryon $i$ we have
\begin{equation}\label{DMS1}
	\Delta M_s(i)=\frac{d_1B_1+d_2B_2}{n_s}.
\end{equation}

The $[70,\ell^+]$ strange resonances of given isospin $I$ and
strangeness
${\cal S}$ $=-n_s$ are shown in Table~\ref{tab1}
together with the expectation values
of $B_1$ and $B_2$ (multiplied by $\sqrt{3}$), the values of
$\Delta M_s(i)$ obtained from
Eq.~(\ref{DMS1}) and the multiplicity $\nu(i)$ of each baryon. The
multiplicity
represents the total number of states of distinct total angular
momentum, obtained from $\ell = 0$, 2,
but having the same values for $I$, ${\cal S}$, $B_1$,
$B_2$ and $\Delta M_s(i)$.

We calculate an average over all strange resonances belonging to
$[56, 2^+]$
or $[70, \ell^+]$ defined as
\begin{equation}\label{ave}
	\Delta M_s =\frac{\sum_i \nu(i)\Delta M_s(i)}{\sum_i \nu(i)}.
\end{equation}
If the average is restricted to members of the $[70,\ell^+]$ multiplet,
by using Table~\ref{tab1} we get
\begin{equation}
\left. \Delta M_s \right|_{[70,\ell^+]} = 77 \pm 61\  {\rm MeV}.
\end{equation}
The total average  including  $[56,2^+]$ with
$\Delta M_s(i) = 206\pm18$~MeV,
$\nu(i) = 18$  and   $[70,\ell^+]$ with
$\Delta M_s(i) = 77 \pm 61$~MeV, $\nu(i) = 36$
is
\begin{equation}
\left. \Delta M_s \right|_{N=2} = 120\pm 47\ {\rm MeV}.
\end{equation}
The error bars on $\Delta M_s(i)$ and on $\Delta M_s$
result from error bars on $d_i$'s. The error bars on $\Delta M_s(i)$
were defined as the quadrature of two uncorrelated errors.

\begin{table}[ht]
\centering
\caption{The strange baryons belonging to the $[70,\ell^+]$ multiplet
with their
isospin $I$ and strangeness ${\cal S}$. Columns 4 and 5 indicate the
expectation
values of the operators $B_1 \sqrt{3}$ and $B_2 \sqrt{3}$ respectively
obtained
from Tables I, II and V of Ref.~\cite{Matagne:2006zf}.
Column 6 gives $\Delta M_s(i)$ obtained from Eq.~(\ref{DMS1}).
Column 7 gives the multiplicity of
the states exhibited in Column 1 (see text).}
\begin{tabular}{ccccccc}
\hline\hline
Baryon & $I$ & ${\cal S}$ & $B_1 \sqrt{3}$ & $B_2\sqrt{3}$ &
$\Delta M_s(i)$ & $\nu(i)$ \\
\hline
$^2\Lambda(70,\ell^+)$ & 0 & $-$1 & $-$1/2 & $-$1 & 64$\pm$58 & 3\\
$^2\Sigma(70,\ell^+)$ & 1 & $-$1 & $-$1/2 & $-$1 & 64$\pm$58 & 3\\
$^2\Xi(70,\ell^+)$ & 1/2 & $-$2 & $-$1 & $-$2 & 64$\pm$58 & 3\\
$^4\Lambda(70,\ell^+)$ & 0 & $-$1 & 0 & $-$3/2 & 254$\pm$47 & 5\\
$^4\Sigma(70,\ell^+)$ & 1 & $-$1 & $-$1 & $-$1/2 & $-$126$\pm$99 & 5\\
$^4\Xi(70,\ell^+)$ & 1/2 & $-$2 & $-$1/2 & $-$5/2&  159$\pm$46 & 5\\
$^2\Sigma'(70,\ell^+)$ & 1 & $-$1 & $-$1/2 & $-$1 & 64$\pm$58 & 3\\
$^2\Xi'(70,\ell^+)$ & 1/2 & $-$2 & $-$1 & $-$2 & 64$\pm$58 & 3\\
$^2\Omega(70,\ell^+)$ & 0 & $-$3 & $-$3/2 & $-$3 & 64$\pm$58 & 3\\
$^2\Lambda'(70,\ell^+)$ & 0 & $-$1 & $-$1/2 & $-$1 & 64$\pm$58 & 3\\
\hline\hline
\end{tabular}
\label{tab1}
\end{table}

\section{Quark model for strange baryons}\label{qmsb}
\subsection{The Hamiltonian}

The potential model used to describe strange baryons is nearly identical
to the
one which was proposed in Ref.~\cite{lnc}. We refer the reader to this
reference
for a detailed discussion of the Hamiltonian, but we nevertheless recall
its main physical content in order to be self-contained.

A baryon, viewed as a bound state of three quarks, can be described in a
first approximation by the following spinless Salpeter Hamiltonian
\begin{equation}\label{ssh}
  H=\sum^3_{i=1}\sqrt{\vec p^{\, 2}_i+m^2_i}+V_Y,
\end{equation}
where $m_i$ is the current mass of the quark $i$, and where $V_Y$ is the
confining interaction potential. Studies based on both the flux tube
model \cite{CKP} and lattice QCD \cite{Koma} suggest that the 
$Y$-junction is
the correct configuration for the flux tubes in baryons: A flux tube,
with energy density (or string tension) $a$, starts
from each quark and the three tubes meet at the Toricelli point of the
triangle
formed by the three quarks. This last point, denoted by $\vec x_{T}$,
minimizes
the sum of the flux tube lengths.
As $\vec x_T$ is a complicated function of the quark positions,
it is useful for our purpose to approximate the genuine confining
potential by the more easily computable expression
\begin{equation}\label{pot1}
V_Y=a \sum^{3}_{i=1}\left|\vec{x}_{i}-\vec{R}\right|,
\end{equation}
where $\vec x_i$ is the position of the quark $i$ is and
$\vec{R}$ the position of the center of mass. The replacement of
the Toricelli point by the center of mass leads to a simplified
confining
potential
which actually overestimates the potential
energy of the genuine $Y$-junction by about 5\% in most cases 
\cite{Bsb04}.
The accuracy of the formula~(\ref{pot1}) is thus rather satisfactory and
can be
improved by a simple rescaling of $a$ \cite{Bsb04}. In Sec.~\ref{su3b},
we shall show how to rescale it correctly. Let us note that,
in Ref.~\cite{lnc}, we used a more complex and accurate approximate form
for
$V_Y$ [see Eq.~(10) of this reference]. But, as we shall see later on,
the inclusion of strange quarks significantly increases the difficulty
of the
analytic work, so that we have to restrict ourselves to the more
tractable
potential~(\ref{pot1}) in order to obtain closed formulas.

Considering only the confining energy is sufficient to understand the
Regge
trajectories of light baryons, but not to reproduce the absolute value
of
their masses. Other contributions are actually needed to lower the mass
spectrum; we shall include them perturbatively. The most widely used
correction
to the Hamiltonian~(\ref{ssh}) is a Coulomb-like interaction of the form
\begin{equation}
  \Delta H_{\textrm{oge}}=-\frac{2}{3}\alpha_S\sum_{i<j}\frac{1}{|
  \vec x_i-\vec x_j|},
\end{equation}
arising from one gluon exchange processes, where $\alpha_S$ is the
strong
coupling constant, usually assumed to be around $0.4$ for light hadrons
\cite{scc,lat0}.

The other interesting contribution to the mass, which can be added
perturbatively as well, is the quark-self energy. Recently, it was shown
that
the quark self-energy, which is created by the color magnetic moment of
a
quark propagating through the vacuum background field, adds a negative
contribution to the hadron masses \cite{qse}. The quark self-energy
contribution for a baryon is given by \cite{qse}
\begin{equation}\label{qsedef}
  \Delta H_{\textrm{qse}}=-\frac{fa}{2\pi}\sum_i\frac{\eta(m_i/\delta)}{
  \mu_i}.
\end{equation}
The factors $f$ and $\delta$ have been computed in lattice QCD studies.
First quenched calculations gave $f = 4$ \cite{qse2}.
A more recent unquenched study gives $f = 3$ \cite{qse3}. Since its
value is
still a matter of research, it may presently be assumed that
$f\in[3,4]$.
Moreover, the value of the gluonic correlation length, denoted as
$\delta$,
is located in the interval $[1.0,1.3]$~GeV \cite{qse2,qse3}.
The function $\eta(\epsilon)$ is analytically known and reads \cite{qse}
\begin{equation}\label{etadef}
\eta(\epsilon)=\left\{
\begin{array}{lll}
\left[\dfrac{-3\epsilon^2}{\left(1-\epsilon^{2}
\right)
^{5/2}}\ln\left(\dfrac{1+\sqrt{1-\epsilon^{2}}}{\epsilon}\right)+\dfrac{
1+
2\epsilon^{2}}{\left(1-\epsilon^{2}\right)^{2}}\right]&(\epsilon < 1)\\
\left[\dfrac{-3\epsilon^2}{\left(\epsilon^{2}-1\right
)^
{5/2}}\arctan\left(\sqrt{\epsilon^{2}-1}\right)+\dfrac{1+2\epsilon^{2}}{
\left(1-\epsilon^{2}\right)^{2}}\right]&(\epsilon > 1) .
\end{array} \right.
\end{equation}
By definition $\eta(0)=1$, and then quickly decreases for higher values
of $\epsilon$, \emph{i.e.} for heavy quarks. It can be checked that, as long as
$\epsilon\lesssim 0.3$, $\eta(\epsilon)$ is approximated with a
reasonable accuracy by
\begin{equation}\label{etaap}
	\hat\eta(\epsilon)=1-\beta\, \epsilon^2
\end{equation}
with $\beta \approx 2.85$. Moreover, this approximation is especially
good for
values of $\epsilon=m/\delta$ corresponding to the strange quark mass
scale.
Consequently, the replacement of $\eta$ by $\hat \eta$ is justified and
it is
sufficient for our purpose. Finally, $\mu_i$ is the dynamical mass of
the quark
$i$, defined as the expectation value \cite{qse}
\begin{equation}\label{muidef}
	\mu_i=\left\langle \sqrt{\vec p^{\, 2}_i+m^2_i}\right\rangle.
\end{equation}
Thus $\mu_i$ is state-dependent, since it is computed by averaging
the kinetic energy of quark $i$ with the wave function of the
unperturbed spinless Salpeter Hamiltonian~(\ref{ssh}).

\subsection{General formulas}\label{genform}

In this work, we are mainly interested in analytical results, needed in
a straightforward comparison with the large $N_c$ mass formula. To this
aim, let
us now introduce auxiliary fields \cite{qse}, in order to get rid of the
square roots appearing in the Hamiltonian~(\ref{ssh}). We obtain
\begin{eqnarray}\label{ham3b}
H(\mu_i,\nu_j)=\sum^3_{j=1}\left[\frac{\vec{p}^{\, 2}_j+m
^2_j}{2\mu_j}+\frac{\mu_j}{2}\right]+\sum^3_{j=1}\left[\frac{ a^2 (\vec{
x}_j-\vec{R})^2}{2\nu_j}+
\frac{\nu_j}{2}\right].
\end{eqnarray}
The auxiliary fields, denoted by $\mu_i$ and $\nu_j$, are, strictly
speaking, operators. Although being formally simpler,
$H(\mu_i,\nu_j)$ is equivalent to $H$ up to the elimination of the
auxiliary fields thanks to the constraints
\begin{subequations}\label{elim}
\begin{eqnarray}
  \delta_{\mu_i}H(\mu_i,\nu_j)&=&0\ \Rightarrow\ \hat \mu_{i}=
  \sqrt{\vec{p}^{\, 2}_i+m^2_i},\\
  \delta_{\nu_j}H(\mu_i,\nu_j)&=&0\ \Rightarrow\ \hat \nu_{i}=
  a|\vec{x}_i-\vec{R}|,
\end{eqnarray}
\end{subequations}
It is worth mentioning that $\left\langle \hat \mu_{i}\right\rangle$ is
nothing
else than the dynamical quark mass introduced in Eq.~(\ref{muidef}), and
that
$\left\langle \hat \nu_{i}\right\rangle$ is the energy of the flux tube
linking
the quark $i$ to the center of mass. Although the auxiliary fields are
operators, the calculations are considerably simplified if one considers
them
as real numbers. They are finally fixed in order to minimize the baryon
mass \cite{Sem03}, and the extremal values of $\mu_i$ and $\nu_j$, 
denoted by $\mu_{i,0}$ and $\nu_{j,0}$, are logically
close to $\left\langle \hat \mu_{i}\right\rangle$ and
$\left \langle \hat \nu_{j}\right\rangle$ respectively.

In Ref.~\cite{coqm}, it has been shown that the eigenvalues of a
Hamiltonian of
the form~(\ref{ham3b}) can be analytically found by
making an appropriate change of variables, the quark coordinates
$\vec x_{i}=\left\{\vec x_{1},\vec x_{2},\vec x_{3}\right\}$ being
replaced by new coordinates
$\vec x^{\, '}_{k}=\left\{\vec R,\vec \xi,\vec \eta\right\}$. The center
of mass is defined as 
\begin{equation}\label{cmdef}
\vec R=\frac{\mu_{1}\vec x_{1}+\mu_{2}\vec x_{2}+\mu_{3}\vec x_{3}}{\mu_
{t}},
\end{equation}
with $\mu_{t}=\mu_{1}+\mu_{2}+\mu_{3}$ and $\{\vec \xi,\vec \eta\}$
being the two relative coordinates. As we consider only light baryons,
composed of
$n$ quarks ($n$ denoting both $u$ or $d$ quarks) and $s$ quarks,
the general formulas obtained in Ref.~\cite{coqm} can be simplified in
the
case where two quarks are of the same mass. Let us set $m_1=m_2=m$. By
symmetry,
we have then $\mu_1=\mu_2=\mu$ and $\nu_1=\nu_2=\nu$. The mass spectrum
of the
Hamiltonian~(\ref{ham3b}) is given in this case by \cite{coqm}
\begin{equation}\label{mass1}	M(\mu,\mu_3,\nu,\nu_3)=\omega_\xi(N_\xi+
3/2)+\omega_\eta(N_\eta+3/2)+\mu+\nu+\frac{\mu_3+\nu_3}{2}+\frac{m^2}{
\mu}+\frac{m^2_3}{2\mu_3},
\end{equation}
where
\begin{equation}
	\omega_\xi=\frac{a}{\sqrt{\mu\nu}},\quad \omega_\eta=\frac{a}{
	\sqrt{2\mu+\mu_3}}\sqrt{\frac{\mu_3}{\mu\nu}+2\frac{\mu}{\mu_3
	\nu_3}}.
\end{equation}
The integers $N_{\xi/\eta}$ are given by
$2n_{\xi/\eta}+\ell_{\xi/\eta}$,
where $n_{\xi/\eta}$ and $\ell_{\xi/\eta}$ are respectively the radial
and orbital quantum numbers relative to the variable $\vec \xi$ or 
$\vec \eta$ respectively. One can also easily check that
\begin{equation}
	\left\langle \vec \xi^{\, 2}\right\rangle=\frac{N_\xi+3/2}{\phi
	\, \omega_\xi},\quad
        \left\langle \vec \eta^{\, 2}\right\rangle=\frac{N_\eta+3/2}{
        \phi\, \omega_\eta},
\end{equation}
with
\begin{equation}
\phi=\sqrt{\frac{\mu^2\mu_3}{2\mu+\mu_3}}	.
\end{equation}
These last identities provide relevant informations about the structure
of the baryons, since
\begin{eqnarray}
	\left\langle \vec X^{\, 2}\right\rangle&=&\left\langle (\vec
	x_1-\vec x_2)^{2}\right\rangle=\sqrt{\frac{4\mu_3}{2\mu+\mu_3}}
	\, \left\langle \vec\xi^{\, 2}\right\rangle,\label{X2def}\\
	 \left\langle \vec Y^{\, 2}\right\rangle&=&\left\langle \left(
	 \frac{\vec x_1+\vec x_2}{2}-\vec x_3\right)^2\right\rangle=
	 \sqrt{\frac{2\mu+\mu_3}{4\mu_3}}\, \left\langle \vec\eta^{\, 2}
	 \right\rangle.\label{Y2def}
\end{eqnarray}
Moreover, by symmetry, we can assume the following equality
\begin{equation}
\left\langle (\vec x_1-\vec x_3)^2\right\rangle	=\left\langle (\vec x_2-
\vec x_3)^2\right\rangle\approx\frac{\left\langle \vec X^{\, 2}\right
\rangle}{4}+  \left\langle \vec Y^{\, 2}\right\rangle,
\end{equation}
which will be useful in the computation of the one gluon exchange
contribution.

The auxiliary fields appearing in the mass formula~(\ref{mass1}) have to
be eliminated by imposing the constraints
$\partial_{\mu_i} M(\mu,\mu_3,\nu,\nu_3)=0$ and
$\partial_{\nu_i} M(\mu,\mu_3,\nu,\nu_3)=0$.
This cannot be done exactly in an analytical way, but, as we shall show
in the
following, solutions can be found by working at the lowest order in
$m^2_i$.
The case $n_s =0$ has been completely treated in Ref.~\cite{lnc}. As
in this last work, we shall assume here that $m_n =0$.

\subsection{The case $n_s=3$}\label{case3s}

Let us begin by the most symmetric case, that is the case of a baryon
formed of three strange quarks (the $\Omega$ family). Then, we have
$m=m_3=m_s$, and
thus $\mu=\mu_3=\mu_s$ and $\nu=\nu_3=\nu_s$ by symmetry.
Equation~(\ref{mass1}) becomes
\begin{equation}
	M(\mu_s,\nu_s)=\frac{a}{\sqrt{\mu_s\nu_s}}(N+3)+\frac{3}{2}\left
	(\mu_s+\nu_s+\frac{m^2_s}{\mu_s}\right),
\end{equation}
where $N=N_\xi+N_\eta$. Because of the symmetry of a $sss$ baryon, its
mass
depends on a single quantum number $N$ only, as for a $nnn$ baryon.
This number $N$ is the total number of excitation quanta associated to
the Hamiltonian~(\ref{ham3b}). It gives the excitation band of the
corresponding eigenstate.

The elimination of $\nu_s$ requires that
\begin{equation}
	\partial_{\nu_s} M(\mu_s,\nu_s)=0\Rightarrow\nu_{s,0}=\left[
	\frac{a^2(N+3)^2}{9\mu_s}\right]^{1/3},
\end{equation}
and then
\begin{equation}\label{m3s1}
	M(\mu_s)=M(\mu_s,\nu_{s,0})=\frac{1}{2}\left[\frac{3^4a^2(N+3)
	^2}{\mu_s}\right]^{1/3}+\frac{3}{2}\left(\mu_s+\frac{m^2_s}{
	\mu_s}\right).
\end{equation}

The constraint $\partial_{\mu_s}M(\mu_s)=0$ does not lead to a tractable
expression for $\mu_s$, unless a development in powers of $m^2_s$ is
performed. One readily finds that
\begin{equation}\label{musdef}
	\mu_{s,0}=\mu_0+\frac{3}{4}\frac{m^2_s}{\mu_0},
\end{equation}
with
\begin{equation}\label{mu0def}
	\mu_0=\sqrt{\frac{a(N+3)}{3}},
\end{equation}
satisfies the relation
$\left.\partial_{\mu_s}M(\mu_s)\right|_{\mu_s=\mu_{s,0}}=0$ at the order
$m^2_s$. Thanks to the relation~(\ref{musdef}), the mass 
formula~(\ref{m3s1}) becomes
\begin{equation}\label{m3s3}
	M=6\mu_0+\frac{3}{2}\frac{m^2_s}{\mu_0}.
\end{equation}

The contributions of the one gluon exchange and of the quark self-energy
can
also be calculated analytically. First, the one gluon exchange mass term
is given by
\begin{equation}
	\Delta M_{\textrm{oge}}=-\frac{2}{3}\alpha_s \sum_{i<j}\left
	\langle \frac{1}{|\vec x_i-\vec x_j|}\right\rangle\approx-\frac{
	2\alpha_s}{\sqrt{\left\langle (\vec x_1-\vec
	x_2)^2\right\rangle}},
\end{equation}
where an obvious symmetry argument has been applied to obtain this last
approximate expression. Equation~(\ref{X2def}) then leads to
\begin{equation}\label{oge3s}
	\Delta M_{\textrm{oge}}=-\frac{2a \alpha_s}{\sqrt 3\mu_0}\left(1
	+\frac{m^2_s}{4\mu^2_0}\right).
\end{equation}
The self-energy term~(\ref{qsedef}), together with the
approximation~(\ref{etaap}), is now given by
\begin{equation}\label{qse3s}
\Delta  M_{\textrm{qse}}=-\frac{3fa}{2\pi}\frac{\hat\eta(m_s/\delta)}{
\mu_{s,0}}
=-\frac{3fa}{2\pi\mu_0}\left[1-\frac{3m^2_s}{4\mu^2_0}-\frac{\beta
m_s^2}{\delta^2}\right].
\end{equation}
The total mass for a triply-strange baryon is finally given by the sum
$M+\Delta M_{\textrm{oge}}+\Delta  M_{\textrm{qse}}$.

It is worth mentioning that in the limit $m_s\rightarrow0$, we recover
the results of Ref.~\cite{lnc}, but the parameter $Q$ of this last
reference has to be set equal to 1 --instead of the
optimal and very close value of 0.93-- in order to take into account our
present assumption that the Toricelli point is located at the center of
mass. When $m_s\neq0$, one can wonder about the validity of the Taylor
expansion in $m^2_s$ that we made. The dominant term of Eq.~(\ref{m3s3})
is $6\mu_0$, while the ``presumably small" term is $3m^2_s/2\mu_0$. In
the worst case, that is for $N=0$, one has
\begin{equation}
\left. \frac{3m^2_s/2\mu_0}{6\mu_0}\right|_{N=0}=\frac{m^2_s}{4a}.
\end{equation}
For typical values $m_s=0.2$~GeV and $a=0.2$~GeV$^2$, this ratio is
around
$0.05$. This justifies a posteriori the relevance of such an expansion.

\subsection{The case $n_s=1$}\label{case1s}

We turn now to the $nns$ baryons. In this case, we can set $m=0$ in the
formula~(\ref{mass1}), and replace the index $3$ by $s$, to make clearly
the appearance of symbols related to the $s$ quark. The mass
formula~(\ref{mass1}) becomes
\begin{equation}\label{mass1s}
M(\mu,\mu_s,\nu,\nu_s)=\frac{\omega_\xi+\omega_\eta}{2}(N+3)+\mu+\nu+
\frac{\mu_s+\nu_s}{2}+\frac{m^2_s}{2\mu_s},
\end{equation}
with
\begin{equation}
	\omega_\xi=\frac{a}{\sqrt{\mu\nu}},\quad \omega_\eta=\frac{a}{
	\sqrt{2\mu+\mu_s}}\sqrt{\frac{\mu_s}{\mu\nu}+2\frac{\mu}{\mu_s
	\nu_s}}.
\end{equation}
An important simplification has been made in Eq.~(\ref{mass1s}):
The term proportional to $\omega_\xi-\omega_\eta$,
present only at $N > 0$ and
vanishing for $nnn$ and $sss$ baryons, was neglected.
This corresponds to the
assumption that the integer $N$ is still a good quantum number to
classify the
asymmetric $nns$ configurations. A numerical resolution of the general
formula~(\ref{mass1}) actually supports this assumption, which is also
made in large $N_c$ QCD. This point will be further investigated in 
Sec.~\ref{Ndep}.

The four auxiliary fields appearing in the mass formula~(\ref{mass1s})
can be eliminated by solving simultaneously the four constraints
\begin{eqnarray}
	\partial_{\mu}M(\mu,\mu_s,\nu,\nu_s)&=&0,\quad 	\partial_{\mu_s}
	M(\mu,\mu_s,\nu,\nu_s)=0,\nonumber\\
		\partial_{\nu}M(\mu,\mu_s,\nu,\nu_s)&=&0,\quad
		\partial_{\nu_s}M(\mu,\mu_s,\nu,\nu_s)=0.
\end{eqnarray}
After some algebra, a solution can be found by working at the order
$m^2_s$,
as we did in the previous section for the case $n_s=3$ (and as we shall
do in
the rest of this paper). In the following, to simplify the notations, we
will write $\mu$ ($\mu_s$) for the optimal
value of the dynamical mass of the $n$ ($s$) quark. We find
\begin{eqnarray}\label{af1s}
	\mu&=&\mu_0+\frac{11}{156}\frac{m^2_s}{\mu_0},\quad \nu=\mu_0+
	\frac{7}{156}\frac{m^2_s}{\mu_0},\nonumber\\
	\mu_s&=&\mu_0+\frac{95}{156}\frac{m^2_s}{\mu_0},\quad \nu_s=
	\mu_0-\frac{53}{156}\frac{m^2_s}{\mu_0},
\end{eqnarray}
where $\mu_0$ is still defined by Eq.~(\ref{mu0def}).
The mass formula~(\ref{mass1s}), in which the auxiliary fields are
replaced by
the expressions~(\ref{af1s}), reads
\begin{equation}
	M=6\mu_0+\frac{1}{2}\frac{m^2_s}{\mu_0}.
\end{equation}

The contribution of the one gluon exchange term is a little more
involved than
in the case $n_s=3$. With the help of relations~(\ref{X2def}) and 
(\ref{Y2def}), it reads
\begin{equation}
	\Delta M_{\textrm{oge}}\approx-\frac{2}{3}\alpha_s
	\left[\frac{1}{\sqrt{\left\langle \vec X^{\, 2}\right\rangle}}+
	\frac{2}{\sqrt{\left\langle \vec X^{\, 2}\right\rangle/4+  \left
	\langle \vec Y^{\, 2}\right\rangle}}\right]=-\frac{2a\alpha_s}{
	\sqrt 3 \mu_0}\left[1+\frac{m^2_s}{12\mu^2_0}\right].
\end{equation}
Relations~(\ref{af1s}) defining $\mu$ and $\mu_s$ allow to write down
the contribution of quark self-energy~(\ref{qsedef}). Using again the
approximation~(\ref{etaap}), we obtain
\begin{equation}\label{qse1s}
\Delta  M_{\textrm{qse}}=-\frac{3fa}{2\pi\mu_0}\left[1-\frac{m^2_s}{4\mu
^2_0}-\frac{\beta m_s^2}{3 \delta^2}\right].
\end{equation}


\subsection{Results for arbitrary $n_s$}

The case $n_s=2$ is very similar to the case $n_s=1$. That is why we
will not treat it explicitly in this paper. Rather, we give here a
summary of the results which are obtained for arbitrary $n_s$. Let us
recall that, when one deals with a baryon made of three massless quarks
($n_s=0$), we recover the results of Ref.~\cite{lnc} with $Q=1$, namely
\begin{equation}\label{mu0def2}
	\mu_0=\sqrt{\frac{a(N+3)}{3}},
\end{equation}
and a total baryon mass, including one gluon exchange and quark self-
energy,
 given by
\begin{equation}\label{mass0s}
	M_0=6\mu_0-\frac{2a\alpha_s}{\sqrt
	3\mu_0}-\frac{3fa}{2\pi\mu_0}.
\end{equation}

By looking at the results of Secs.~\ref{case3s} and \ref{case1s}, one
can
deduce that the auxiliary fields $\mu$ and $\mu_s$ have the following
general form
\begin{eqnarray}
	\mu=\mu_0+\frac{11\, n_s}{156}\frac{m^2_s}{\mu_0},\quad\quad &
	n_s=0,1,2, \\
	\mu_s=\mu_0+\frac{84+11\,  n_s}{156}\ \frac{m^2_s}{\mu_0},\quad
	\quad &n_s=1,2,3.
\end{eqnarray}
Moreover, the total baryon mass is given by
\begin{equation}\label{massgene}
	M=M_0+n_s\, \Delta M_s,\quad\quad n_s=0,1,2,3,
\end{equation}
where the contribution of the $s$ quarks is
\begin{equation}\label{dmns}
	\Delta M_s=\left[\frac{1}{2}-\frac{\alpha_s a}{6\sqrt 3 \mu_0^2}
	+\frac{f a}{2\pi}\left(\frac{3}{4\mu^2_0}+
	\frac{\beta}{\delta^2}\right) \right] \frac{m^2_s}{\mu_0}.
\end{equation}
These formulas are only valid at the order $m^2_s$. We checked that they
agree
with the explicit calculation in the case $n_s=2$.
So each $s$ quark brings the same contribution to the baryon mass and
this contribution depends on $m_s$.

The eigenvalues of the spinless Salpeter Hamiltonian with the
potential~(\ref{pot1}) have been numerically calculated in order to
check
the accuracy of the mass formula~(\ref{massgene}) with
$\alpha_s = f = 0$ (one gluon exchange and self-energy are treated as
perturbations). For the relevant values of $m/\sqrt{a}$ and $N$,
the relative error is found around 10\%.

\section{Comparison of the two approaches}\label{compar}

\subsection{SU(3)-breaking mass term}\label{su3b}

In both the $1/N_c$ expansion and the quark models the baryon mass
is affected by an explicit SU(3)-breaking due to the mass difference
between nonstrange $u,d$ and strange $s$ quarks. The effect of
SU(3)-breaking in the $1/N_c$ expansion  mass formula has been estimated
in Sec.~\ref{s_bar} through terms including the operators $B_i$.
Obviously,
a nonvanishing value of the strange quark mass also requires the quark
model mass formula to be modified as in Eq.~(\ref{massgene}).
Then, the corresponding SU(3)-breaking mass terms~(\ref{dmns}) can be
compared
to those resulting from Eq.~(\ref{break}). To do this, we have to
determine the values of the parameters in the quark model,
since the coefficients of the large $N_c$ formula have already been
fitted on the experimental data.

First, we recall that the auxiliary field method yields upper bounds of
the mass
spectrum, as it is shown in Ref.~\cite{hyb1}. This artifact can be cured
by
making a rescaling of the string tension $a$, so that the Regge slope of
$nnn$ baryons is equal to the Regge slope of $n \bar n$ mesons
\cite{lnc}.
Obtained within the flux tube model, this slope is $2\pi\sigma$,
$\sigma$ being
the physical string tension. By looking at the formula~(\ref{mass0s}),
we see
that a correct rescaling is made by taking $a=\pi\sigma/6$.
In Ref.~\cite{lnc}, we have shown that a remarkable compatibility
between large
$N_c$ QCD  and quark model results exists for the nonstrange baryon
masses,
provided we take $\sigma=0.163$~GeV$^2$, $\alpha_s=0.4$, and $f=3.5$.
These are also the values considered in this work,
despite the fact that these parameters were obtained with a value
$Q=0.93$
instead of the value $Q=1$ assumed here (see Sec.~\ref{case3s}).
However, two extra parameters are
present when strange quarks are taken into account. These are $m_s$ and
$\delta$. The value $\delta=1.0$~GeV has already been used in
potential models for mesons, in good agreement with the experimental
data
\cite{qse,expe}. We shall thus use it here too. Finally, $m_s$ was
fitted to get an optimal
agreement between the quark model and the large $N_c$ QCD mass shift at $N$ = 0.
We actually found $m_s=0.243$~GeV, which is larger than the PDG
value of
95$\pm$25~MeV \cite{PDG}. However, a strange quark mass in the range
$0.2$-$0.3$~GeV is quite usual in potential models \cite{Lucha}. All
parameters are gathered in Table~\ref{parmod}.

Following the error analysis of Ref.~\cite{lnc}, 
$\sigma=0.163\pm 0.004$~GeV$^2$,
$\alpha_s=0.4\pm 0.05$, and $f=3.5\pm 0.12$. We know that
$\delta \in [1.0,1.3]$~GeV. If we allow a variation of 10\% for the
parameter $\beta$, we find an error on $m_s$ around 12~MeV.

\begin{table}[ht]
\centering
\caption{Parameters of the model.}
\begin{tabular}{ll}
\hline\hline
$a=\pi \sigma /6$ & $\delta=1.0$~GeV \\
$\sigma=0.163$~GeV$^2$ & $\beta=2.85$ \\
$\alpha_s=0.4$ & $m_n=0$ \\
$f=3.5$ &  $m_s=0.243$~GeV \\
\hline\hline
\end{tabular}
\label{parmod}
\end{table}

A comparison between the mass shift $\Delta M_s$, obtained with the
quark model and its large $N_c$ counterpart,
is given in Table~\ref{tabdm} for $N = 0, 1, 2, 3, 4$. One can see that
the quark model predictions are always
located within the error bars of the large $N_c$ results.
Except for $N = 3$, the central values of $\Delta M_s$ in the large
$N_c$ approach are close to the quark model results.
Ignoring the large $N_c$ value at $N = 3$, which would require
further investigations, as we argued in Sec.~\ref{mass_f}, 
one can see that $\Delta M_s$ 
decreases slowly and 
monotonously with increasing $N$, in both methods. This
suggests that
the central value of $\Delta M_s$ obtained in Ref.~\cite{GM} in the
$1/N_c$ approach is probably far too small for $N=3$.

\begin{figure}[ht]
\includegraphics*[width=9cm]{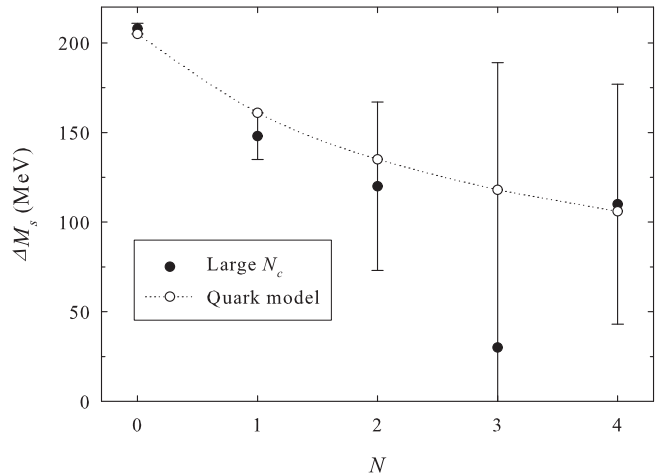}
\caption{Plot of the results presented in Table~\ref{tabdm}. The quark
model predictions for the mass shift $\Delta M_s$ (empty circles) are compared
to the large $N_c$ data (full circles). A dotted line links the quark model
points to guide the eye.}
\label{Fig1}
\end{figure}

The results of
Table~\ref{tabdm} are plotted in Fig.~\ref{Fig1} to see more clearly the
evolution of the mass shifts with $N$. Thus, in both approaches, one
predicts a
mass shift correction term due to SU(3)-breaking which decreases with
the excitation energy (or $N$).

\begin{table}[ht]
\centering
\caption{Mass shifts $\Delta M_s$ (MeV) given by Eq.~(\ref{dmns})
with the parameters of Table~\ref{parmod} for the quark model, compared
to large $N_c$ mass shifts
for various values of $N$: $N=0,1,3$  from Ref.~\cite{GM}, $N$ = 4
from Ref.~\cite{Matagne:2004pm}; the $N=2$ case is studied in detail
in Sec.~\ref{lnc_s}.}
\begin{tabular}{ccc}
\hline\hline
$N$ &  Quark model   & Large $N_c$  \\
\hline
0 & 205 & 208$\pm$3 \\
1 & 161 & 148$\pm$13\\
2 & 135 & 120$\pm$47 \\
3 & 118 &  30$\pm$159\\
4 & 106 & 110$\pm$67\\
\hline\hline
\end{tabular}
\label{tabdm}
\end{table}

\subsection{The dependence on $N$}\label{Ndep}

When $n_s=0$ or 3, the symmetry of the problem leads to a mass formula
which depends on $N=N_\xi+N_\eta$ only,
with $N_\xi$ and $N_\eta$ introduced in Sec.~\ref{genform}. When $n_s=1$
or 2 however, this is
not the case. In order to perform explicit calculations, we have
assumed that $N$ is still a good quantum
number to classify baryon states with one or two strange quarks. This
assumption also
ensures that the total parity remains $(-1)^N$ in a given band.

To quantitatively support the above considerations, we can now build
a quantity to estimate the validity of this approximation for $n_s=1$ or 2.
In this case, the
general mass formula~(\ref{mass1}) must be used. It depends on the
auxiliary fields, that we commonly denote here as
$\varphi_i$, but also on $N_\xi$ and $N_\eta$. The value of
$M(N_\xi,N_\eta,\varphi_i)$, thus given by Eq.~(\ref{mass1}), can be
computed once $\sigma$ and $m_s$ are fixed. We choose
$\sigma=0.163$~GeV$^2$ and $m_s=0.243$~GeV as in the previous section.
First, instead of $N_\xi$ and $N_\eta$, we work with the quantum numbers
$N$ and $N'=N_\xi$, that is to say with the mass formula
$M(N',N-N',\varphi_i)$
where $N'=0,\dots,\, N$. Once $N$ and $N'$ are fixed, standard numerical
routines allow to minimize the mass with respect to the auxiliary
fields.
This leads to the optimal values $\varphi_{i,0}$ and finally to
$M(N',N-N',\varphi_{i,0})$. Then, we define
\begin{equation}
\delta M(N)=\frac{\max\{M(N',N-N',\varphi_{i,0})\}-
\min\{M(N',N-N',\varphi_{i,0})\}}{\left[\sum^N_{N'=0}M(N',N-N',
\varphi_{i,0})\right]/\left[N+1\right]},
\end{equation}
where the maximal and minimal masses are chosen within the set of
allowed $N'$ for a given $N$.
$\delta M$, which  depends only on
$m_s/\sqrt{\sigma}$, is a measure of the quality of $N$ as a good
quantum number: The more $\delta M$ is small, the less the value of the
mass at a given $N$ depends on the other quantum number $N'$.

\begin{figure}[ht]
\includegraphics*[width=9cm]{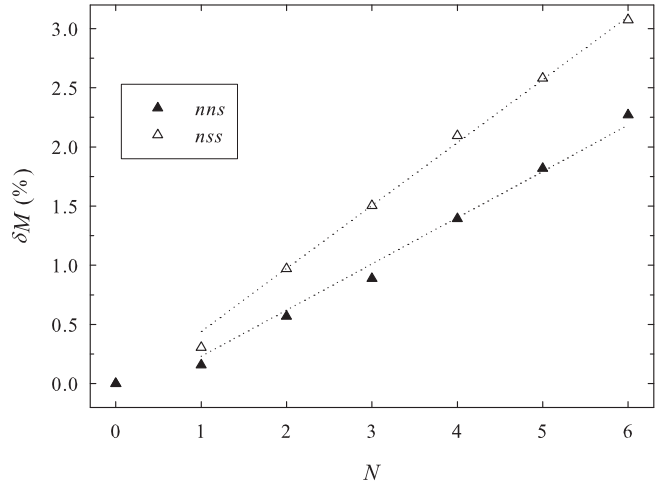}
\caption{Plot of $\delta M$ ($\%$) versus $N$ for $nns$ baryons
(full triangles) and for $nss$ baryons (empty triangles), with the parameters 
of Table~\ref{parmod}.
Linear fits for the $N>0$ points are plotted with a dotted line.}
\label{Fig2}
\end{figure}

A plot of $\delta M$ versus $N$ is presented in Fig.~\ref{Fig2} for the
$nns$ and $nss$ baryons. By definition, $\delta M(0)=0$ since the only
possibility
is $N'=0$ in this case. Then, it appears that $\delta M$ increases
linearly for $N\geq1$. Moreover, the values vary very slowly with
$m_s/\sqrt{\sigma}$. The key point to observe in this graph is that
$\delta M(N\leq 6)\lesssim$ 3\%.
As no experimental state such that $N>6$ has  so far been observed, we
can conclude that the mass formula obtained from the spinless Salpeter
Hamiltonian~(\ref{ssh}) mainly depends on $N$: A change of $N'$ at a
given
$N$ only causes a change of the mass which is lower than 3\% in all the
case that are relevant with regard to current experimental data.

Let us note that the previous result is only strictly valid for the mass
formula obtained from the Hamiltonian~(\ref{ham3b}). When the
eigenstates of
the Hamiltonian~(\ref{ssh}) are computed in harmonic oscillator bases,
it can
be seen that these eigenstates contain components from different $N$
bands;
thus $N$ is only an approximate good quantum number. Nevertheless,
the band mixing is usually small and changes by less than 10\% the mass
of a state with a dominant component in a given band $N$.
\subsection{Regge trajectories}

Since $N$ appears to be a relevant quantum number to classify light
baryons with a good accuracy, it is
of interest to study the predictions of large $N_c$ QCD and quark model
regarding the Regge trajectories of nonstrange as well as of strange
baryons. At the leading order in $N$, we actually expect that $M^2\propto N$.
Indeed, formula~(\ref{massgene}) tells us that, at large $N$,
\begin{eqnarray}\label{msqr}
M^2&\approx &M^2_0+2 M_0 n_s\Delta M_s\nonumber\\
&=&2\pi\sigma(N+3)-\frac{4}{\sqrt 3}\pi\sigma\alpha_s-3f\sigma+\left[6+
\frac{f\sigma\beta}{\delta^2}\right]n_s m_s^2.
\end{eqnarray}
Our particular quark model thus states that baryons should follow Regge
trajectories with a common slope, irrespective of the strangeness of the
baryons. This feature has also been pointed out in other approaches
based on
the diquark-quark picture \cite{simobar,SW}. In Ref.~\cite{lnc} we have
already shown that, for nonstrange baryons, $M^2_0$ was actually equal
to $(N_c\, c_1)^2$, this last quantity being the dominant term in the
large $N_c$ mass formula. Moreover, following Ref.~\cite{GM}, the fitted
values of $c_1$ does not change whether or not strange quarks are taken
into account. Therefore, the Regge slope of strange and nonstrange baryons is
also predicted to be independent of the strangeness in the $1/N_c$
expansion method.

However, the intercept depends on the number of strange quarks.
Following the
explicit formula~(\ref{msqr}), it logically increases for larger values
of $n_s$ and $m_s$. Formally, the contribution of strange quarks to the
intercept
is given in the quark model by $\left.	2 M_0 \Delta M_s\right|_{N=0}$,
and in the large $1/N_c$ expansion by
$2\, N_c\, c_1\, \left.\Delta M_s \right|_{N=0}$.
Taking the values of $M_0$ and $c_1$ from Ref.~\cite{lnc}, and the
values of $\Delta M_s$ found in this paper, both large $N_c$ QCD and
quark model agree on the value of the intercept. The first method leads
to $0.361\pm0.005$~GeV$^2$, while the second one gives $0.355$~GeV$^2$.

The light baryon Regge trajectories are thus predicted to share a common
slope, but we expect that they should be separated into parallel straight lines
with an intercept depending on the strangeness. Unfortunately, too few
experimental data are currently known at large excitation energies (large $N$) to
check this picture. But, it could be used as an interesting tool to identify
strange and nonstrange excited baryons in future experiments.

%
\section{Conclusion}\label{conclu}

The previous work establishing a connection between the quark model and
the $1/N_c$ expansion method has been successfully extended to
include strange baryons with nonzero mass $m_s$. A comparison between
the SU(3)-breaking terms in the mass formula of the two approaches has been
made and we found a good quantitative agreement. The
comparison was possible through the introduction of a band
quantum number $N$, customarily used in the baryon classification.
While for nonstrange baryons $N$ appears straightforwardly, the
inclusion of strange quarks with nonzero mass turned out to be more elaborate.
However we have  numerically proved that $N$ can be considered as a good quantum
number in a realistic quark model with a $Y$-junction confinement by
keeping terms up to order $m^2_s$ in the Taylor expansion. 

\acknowledgments
Financial support is acknowledged by C.~S. and F.~B. from FNRS
(Belgium).

\end{document}